\documentclass{article}
\usepackage[utf8]{inputenc}
\usepackage{graphicx}
\usepackage{wrapfig}
\usepackage{lipsum}
\usepackage{physics}
\usepackage{caption}
\usepackage{subcaption}
\usepackage{setspace}
\usepackage{authblk}
\usepackage[hyphens]{url}
\usepackage{hyperref}
\usepackage[nottoc]{tocbibind}
\usepackage{float}
\usepackage{mathtools}
\usepackage{subcaption}
\usepackage{color}
\usepackage[dvipsnames]{xcolor}
\usepackage{breqn}
\usepackage{url}
\usepackage{verbatim}
\usepackage{amsmath}
\usepackage{multicol}
\usepackage[margin=0.9in]{geometry}
\usepackage[labelfont=bf,labelsep=space,font=footnotesize]{caption}

\title{\textbf{\Large{Capturing electronic correlations in electron-phonon interactions in molecular systems with the $GW$ approximation}}}

\author[1,2,*]{Antonios M. Alvertis}
\author[3]{David B. Williams-Young}
\author[4]{Fabien Bruneval}
\author[2,5,6,$\dagger$]{Jeffrey B. Neaton}

\affil[1]{KBR, Inc., NASA Ames Research Center, Moffett Field, California 94035, United States}
\affil[2]{Materials Sciences Division, Lawrence Berkeley National Laboratory, Berkeley, California 94720, United States}
\affil[3]{Applied Mathematics and Computational Research Division, Lawrence Berkeley National Laboratory, Berkeley, California 94720, United States}
\affil[4]{Universit\'{e} Paris-Saclay, CEA, Service de Corrosion et de Comportement des Mat\'{e}riaux, SRMP, 91191 Gif-sur-Yvette, France}
\affil[5]{Department of Physics, University of California Berkeley, Berkeley, United States}
\affil[6]{Kavli Energy NanoScience Institute at Berkeley, Berkeley, United States}
\affil[*]{e-mail: amalvertis@lbl.gov}
\affil[$\dagger$]{e-mail: jbneaton@lbl.gov}


\date{}

\begin{document}

\maketitle

\noindent
\section*{Abstract}

\noindent
Electron-phonon interactions are of great importance
to a variety of physical phenomena, and their accurate
description is an important
goal for first-principles calculations. Isolated examples of materials and molecular systems have 
emerged where
electron-phonon coupling is enhanced over density functional theory (DFT)
when using the Green's-function-based \emph{ab initio} $GW$ method, which provides a more accurate description of electronic correlations. 
It is however unclear how 
general this enhancement is, and how employing high-end quantum chemistry
methods, which further improve the description of electronic correlations, might further alter electron-phonon interactions over $GW$ or DFT.
Here, we address these questions by computing the renormalization of the highest occupied molecular orbital energies of Thiel's set of organic molecules by harmonic vibrations using DFT, $GW$ and equation-of-motion coupled-cluster calculations. We find that  $GW$ can increase the magnitude of the
electron-phonon coupling across this set
of molecules by an average factor of $1.1-1.8$ compared to DFT, while equation-of-motion coupled-cluster leads to an increase of $1.4-2$. The electron-phonon coupling predicted with the \emph{ab initio} $GW$ method
is generally in much closer agreement to coupled cluster values compared to DFT, establishing $GW$ as an accurate way of computing electron-phonon phenomena in molecules and beyond at a much lower computational cost
than higher-end quantum chemistry techniques.

\clearpage



\section{Introduction}
\label{introduction}
The interaction of electrons with atomic and lattice
vibrations (phonons) in solids is of central 
importance in chemistry and physics, defining carrier
transport properties~\cite{OReilly2014,Franke2012}, superconducting critical
temperatures~\cite{McMillan1968,Allen1975}, the rate of non-radiative 
recombination~\cite{Shi2012,Shi2015}, the reaction pathway of ultrafast charge and energy transfer~\cite{Sukegawa2014,Schnedermann2019}, and more for materials and molecular systems. In organic 
molecules and molecular crystals, atomic and lattice vibrations
can cause substantial renormalization of ground 
and excited
state properties~\cite{Hele2021,Alvertis2020,Brown-Altvater2020,Kapil2022}. Therefore, accurate and
predictive theories for the interaction of electrons
with vibrations are important in order to
guide the experimental search for systems with a wide range of functionality, as well as to promote
the fundamental understanding of a variety of 
physical processes.

For decades, density functional theory (DFT)
has arguably been the most widely used framework for computing
the properties of molecules and materials from first principles, and it has been extensively used to calculate
electronic and vibrational degrees of freedom, as well as their interplay~\cite{Fan1992,Wong1996,Heinemeyer2008,Casula2011,Shang2021}. The choice of
the DFT exchange-correlation functional has been known
in several systems to affect the predicted magnitude of these
effects, with examples demonstrating that
inclusion of exact exchange to the DFT functional
tends to increase electron-phonon coupling~\cite{LaflammeJanssen2010} compared to cases where the local density approximation (LDA)~\cite{Jones1989} and generalized gradient approximations (GGA)~\cite{pbe} are used. Moreover, \emph{ab initio} many-body perturbation theory within the $GW$ approximation~\cite{Lars1965,Hybertsen1986}, which is known to yield highly accurate properties for the electronic levels of molecular systems and solids~\cite{shirley_prb1993,grossman_prl2001,rostgaard_prb2010,Blase2011,bruneval_jcp2012,ren_njp2012,sharifzadeh_epjb2012,korzdorfer_prb2012,Bruneval2013,vansetten_jctc2013,koval_prb2014,govoni_jctc2015,vansetten_jctc2015,knight_jctc2016,kawahara_prb2016,blase_jcp2016,maggio_jctc2017b,lange_jctc2018,golze_jctc2018,wilhelm_jcpl2018,lewis_jctc2019,golze_fchem2019,Bruneval2021}, has also been
used to compute electron-phonon coupling, showing significantly increased
magnitude of these interactions over DFT in some cases~\cite{Faber2011,Yin2013,Antonius2014,Monserrat2016,Li2019}, but only small changes in others~\cite{Monserrat2016}. Further increases of electron-phonon coupling might also be expected with an even more accurate description of electronic correlations, \emph{e.g.} methods such as coupled cluster (CC)~\cite{Bartlett2007}. 
Additionally, the challenge associated with rigorously converging electron-phonon calculations in solids has led
to the suggestion that the discrepancies reported in the
literature in the phonon-induced fundamental gap renormalization obtained at different levels of electronic structure theory could potentially be attributed to under-converged calculations~\cite{Miglio2020}. Given these practical challenges and inconclusive reports, 
there remains no consensus on whether including
exact exchange within DFT or employing higher-level $GW$ calculations leads to a systematic increase
of the computed electron-phonon interactions, or
whether examples in the literature showing such an increase constitute isolated cases. 
To complicate matters further, Kohn-Sham wavefunctions computed within DFT are most commonly used as a starting point for $GW$ calculations, 
and the choice of starting point can greatly affect the $GW$ results~\cite{Caruso2016,Kaplan2016}. However, the dependence of
the electron-phonon interactions computed with $GW$ on the functional employed for the DFT starting point remains unknown. 

Here we present a systematic study of electron-phonon coupling using Thiel's set of organic 
molecules~\cite{Schreiber2008}, a set of 28 small
molecules. As a proxy for the magnitude of the coupling of molecular vibrations to electrons,
we study the renormalization of the highest occupied molecular orbital (HOMO) energy of each molecule due to the zero-point motion of the atomic nuclei, which we obtain within the harmonic approximation using a well-established finite-displacements approach~\cite{Monserrat2018,Hele2021}. We compute the zero-point renormalization (ZPR) of the HOMO energies with DFT using several different functionals with varying degrees of exact exchange, and we also perform $GW$ calculations with various
DFT starting points. We compare our DFT and $GW$ ZPR calculations to the values obtained with accurate coupled-cluster (CC) methods. 
Our results
establish that the correction to DFT electronic correlations obtained within the $GW$ and CC formalisms tends to systematically increase the magnitude of
electron-phonon coupling, particularly when employing DFT functionals with semilocal or small amounts of
exact exchange. That $GW$ and CC
methods generally predict stronger electron-phonon interactions compared to DFT can have implications for accurate prediction of properties such as superconductivity~\cite{Yin2013,Li2019} and
finite-temperature excited state properties~\cite{Alvertis2023} in other systems. The results with $GW$
are in overall good agreement with CC, yet obtained at
much lower computational cost ($N^4$ compared to $N^6$ for coupled cluster), suggesting that the $GW$ method provides a powerful and computationally efficient way of accurately modeling
electron-phonon interactions in molecular systems.

\section{Theoretical background and computational methods}
\label{background}

\begin{figure}[tb]
    \centering
    \includegraphics[width=0.6\linewidth]{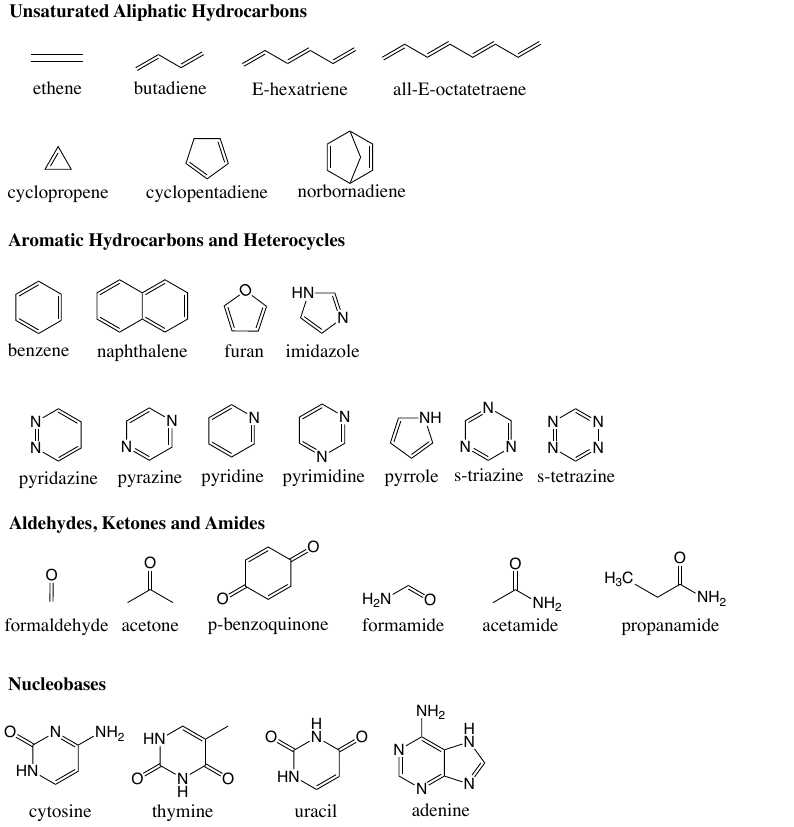}
    \caption{Thiel's set of organic molecules.}
    \label{fig:structures}
\end{figure}

In this paper we study the so-called Thiel's set of organic molecules shown in Fig.\,\ref{fig:structures}, which
consists of $28$ diverse gas-phase molecular
structures~\cite{Schreiber2008}. In order to understand the impact of the level of the
electronic structure theory on the
magnitude of the electron-phonon interactions, we focus on the
vibrationally-induced ZPR of the HOMO energy of the Thiel's set
molecules. Section\,\ref{vibrational_averages} provides an overview
of the theoretical framework we employ here. Then Section\,
\ref{electronic_structure} goes into some detail on the electronic
structure methods used in this work in order to compute the HOMO energies of the Thiel's set
molecules, and Section\,\ref{ZPR} summarizes the results of our
vibrational and electronic structure calculations to compute the HOMO ZPR at the
various levels of theory. 

\subsection{Vibrational averages of observables}
\label{vibrational_averages}
In what follows, we review the formalism for computing vibrational averages for quantum-mechanical operators, and some of the approximations relevant to this work. Let us consider an operator $\mathcal{O}$ corresponding to an observable of interest, in the presence
of atomic motion at temperature $T$. 
Within the Born-Oppenheimer approximation, the finite-temperature expectation value of $\mathcal{O}$ may be written
as
\begin{equation}
    \left\langle \mathcal{O}(T) \right\rangle_{\cal H}
    = \frac{1}{Z} \int dX \mathcal{O}(X) e^{ -\beta \cal H },
    \label{eq:ens_avg}
\end{equation} 
where the canonical partition function $Z = \int dX e^{ -\beta \cal H }$ involves the configuration space integral $\int dX$~\cite{Patrick2014}.

The Hamiltonian $\cal H$ of the system includes electronic and nuclear degrees of freedom in general, and may be approximated
at different levels. One approach that is generally valid for small organic molecules is to assume nuclear motion to be harmonic, reducing the vibrational contribution to the Hamiltonian to the following form,
\begin{equation}
    {\cal H}^{\textrm{har}} 
    \equiv 
    \frac{1}{2} \sum_{n}( \nabla_{u_{n}}^2 +
    \omega_{n}^2 u_{n}^2),
\label{eq:har_H}
\end{equation}
in atomic units. Here, $\omega_n$ is
the frequency of the $n_{\text{th}}$ vibrational mode, and $u_n$ the respective eigendisplacement, which are computed using a finite-displacements approach~\cite{Kresse1995, Parlinski1997} in conjunction with DFT total energy and forces calculations. The ground state vibrational properties
of organic molecules are known to be fairly insensitive to the amount
of exact exchange in the employed DFT functional~\cite{LaflammeJanssen2010}, and we therefore
compute the vibrational modes and frequencies of all structures considered here with the global hybrid B3LYP~\cite{Becke1993}, as implemented
in the NWChem software~\cite{Apra2020}. 

In the harmonic approximation, the
integral of eq.\,\ref{eq:ens_avg} becomes~\cite{Monserrat2018}
\begin{equation}
    \left\langle \mathcal{O}(T) \right\rangle_{\cal H}=\int d\mathbf{u}|\Phi(\mathbf{u};T)|^2\mathcal{O}(\mathbf{u}), \label{eq:harm_exp_value}
\end{equation}
where $\mathbf{u}$ a vector of atomic displacements associated with a given vibrational mode
in the system and 
\begin{equation}
    |\Phi(\mathbf{u};T)|^2 = \prod_{n}(2\pi \sigma^2_{n}(T))^{-1/2}\exp{\left(-\frac{u_{n}^2}{2\sigma^2_{n}(T)}\right)}
    \label{eq:harm_density}
\end{equation}
is the harmonic density at temperature $T$, which in turn is a product of Gaussian functions of width,
\begin{equation}
    \sigma^2_{n}(T) = \frac{1}{2\omega_{n}}\cdot \coth{\left(\frac{\omega_{n}}{2k_BT}\right)}.
    \label{eq:Gaussian_width}
\end{equation}
The expectation value of eq.\,\ref{eq:harm_exp_value} may be computed by a Monte Carlo sampling, drawing $N$
random samples $\mathbf{u}_i$ from the harmonic density distribution function of eq.\,\ref{eq:harm_density}, at which
we compute the observable of interest $\mathcal{O}(\mathbf{u}_i)$,
allowing us to finally obtain
\begin{equation}
    \left\langle \mathcal{O}(T) \right\rangle_{\text{MC}}
    = 
    \lim_{N \rightarrow \infty} \frac{1}{N}\sum_{i=1}^{N} \mathcal{O}(\mathbf{u}_i).
    \label{eq:har_band_gap}
\end{equation}
This Monte Carlo (MC) sampling of the
expectation value of observables has
strong parallels to the nuclear ensemble method~\cite{Crespo-Otero2012}. 

Apart from the harmonic and adiabatic
approximations, no further assumptions
have been made in deriving eq.\,\ref{eq:har_band_gap}. The Monte
Carlo sampling method has the advantage
of including the effect of vibrations
to all orders on the observable of
interest in a non-perturbative fashion~\cite{Monserrat2015}.
We will focus here on the effect of vibrations
on electronic observables (see Section\,\ref{electronic_structure}) and refer to the interactions between electronic and vibrational
degrees of freedom as electron-phonon interactions, 
and use the term ``phonon" interchangeably with
``molecular vibrations" and ``normal modes".

The Monte Carlo sampling method does not allow one
to identify the contribution of individual 
vibrational modes to the thermal
average, as we always consider collective displacements. In order
to isolate the effect of individual
vibrations one can expand the observable
of interest in a specific vibrational coordinate as
\begin{equation}
    \label{eq:quadratic_expansion}
    \mathcal{O}(\mathbf{u}) = \mathcal{O}(\mathbf{0})+\sum_{n} \frac{\partial \mathcal{O}(\mathbf{0})}{\partial u_{n}}u_{n}+\frac{1}{2}\sum_{n}\sum_{n'}\frac{\partial^2 \mathcal{O}(\mathbf{0})}{\partial u_{n}\partial u_{n'}}u_{n}u_{n'}+... \hspace{0.1cm},
\end{equation}
where $\mathbf{u}=\mathbf{0}$ represents
the equilibrium geometry of the molecule. In the above expansion of $\mathcal{O}(\mathbf{u})$ the third-order term vanishes, hence by truncating eq.\,\ref{eq:harm_exp_value} to fourth order one arrives at the so-called quadratic (Q) expectation value:
\begin{equation}
    \label{eq:quadratic}
    \left\langle \mathcal{O}(T) \right\rangle_{\text{Q}} = \mathcal{O}(\mathbf{0})+\sum_{n} \frac{1}{2\omega_{n}}\cdot \frac{\partial^2\mathcal{O}}{\partial u_{n}^2}[\frac{1}{2}+n_B(\omega_{n},T)],
\end{equation}
where $n_B(\omega_{n},T)$ is the Bose-Einstein distribution function
for the occupation of mode $n$ at temperature $T$. At a practical level, the expectation value of eq.\,\ref{eq:quadratic} may be obtained by approximating the second derivative appearing for every vibrational mode in this
equation by the finite-difference formula
\begin{equation}
    \label{eq:second_der}
    \frac{\partial^2\mathcal{O}}{\partial u_{n}^2} \approx \frac{\mathcal{O}(\delta u_{n})+\mathcal{O}(-\delta u_{n})-2\mathcal{O}(\mathbf{0})}{\delta u_{n}^2}.
\end{equation}
The quadratic expectation value of eq.\,\ref{eq:quadratic} is less
accurate than the Monte Carlo expectation value of eq.\,\ref{eq:har_band_gap}, and also has the disadvantage that the 
finite difference formula employed to approximate the second derivative introduces some dependence on the choice of $\delta u_n$ for every mode. While we will not use this level of approximation
to extract quantitative values of observables in the presence of
molecular vibrations, it offers the advantage of separating the
contribution of each normal mode, and we will employ it below to gain
further physical insights into the ZPR we compute with eq.\,\ref{eq:har_band_gap}.

\subsection{Electronic structure calculation of molecular ionization potentials}
\label{electronic_structure}
Here we are interested in understanding
the effect of vibrations on the HOMO energies, $\epsilon_{\text{HOMO}}$, of the Thiel's set molecules, therefore we
take $\mathcal{O}=\epsilon_{\text{HOMO}}$, using the notation of Section\,\ref{vibrational_averages}.
We compute all HOMO energies with
DFT using a range of exchange-correlation functionals. Specifically, 
we employ the GGA as formulated by Perdew and Burke and Ernzerhof (PBE)~\cite{pbe}; and we employ
global hybrid variants of PBE, namely PBEh($\alpha$), that contain a finite amount of
exact exchange that is governed by the parameter $\alpha$ ranging from $0$ to $1$, namely
\begin{equation}
    \label{eq:PBEh}
    E_{XC}=\alpha E^{EX}_X+(1-\alpha)E^{PBE}_X+E^{PBE}_C,
\end{equation}
where $E_{XC}$ the PBEh exchange-correlation functional, $E^{EX}_X$ the exact exchange energy, and $E^{PBE}_X$ and $E^{PBE}_C$ are the PBE exchange and correlation energies respectively. We consider two cases, $\alpha=0.25$ and $\alpha=0.5$, in order to gain insights into
the impact of the amount of exact exchange on the HOMO energies and their ZPR in this set of molecules. The case with $\alpha=0.25$ is also referred to as the PBE0 functional, and we will refer to the case with $\alpha=0.5$ simply as the
PBEh functional. Moreover, we employ
the BHLYP functional~\cite{10.1063/1.464304}, which also contains $50\%$ exact exchange. 
Functionals with about 50~\%  are known to provide an excellent starting point for subsequent $GW$ calculations~\cite{Bruneval2013,Bruneval2021}
for molecular systems. 

Beyond DFT, we also perform $GW$ calculations using the so-called one-shot $G_oW_o$ approach. Here the one-particle Green's functions $G$ and screened Coulomb interaction $W$ are constructed
from the eigenvalues and eigenvectors
of a preceding DFT calculation ($W$ is obtained within the random phase approximation~\cite{Ceperley1980}), and used to construct the self-energy $\Sigma=iGW$, which in turn is used to generate
quasiparticle energies via correction $\Sigma-V_{XC}$ to generalized Kohn-Sham eigenvalues. Within this
one-shot approach, we do not iteratively update $\Sigma$ based on
the computed $GW$ eigenvalues and eigenvectors, as the one-shot $G_oW_o$ approach has been shown to produce highly accurate results for a wide range of molecular systems~\cite{Bruneval2021}.
While larger basis sets are often required to achieve convergence of the orbital energies obtained within \emph{ab initio} many-body perturbation theory~\cite{Bruneval2020}, the ZPR is computed as a difference between orbital energies (see eq.~\ref{eq:ZPR} below), which converges faster. In the supplementary information Fig.~1 and Fig.~2 we plot the convergence of the HOMO energy and HOMO ZPR respectively, for the example case of pyrimidine, with respect to the basis set size. We thus verify that the aug-cc-pVTZ basis produces converged results for the HOMO energy ZPR, which are at most $1$\,meV different from the values obtained using aug-cc-pV5Z.


In order to assess the accuracy of
the computed DFT and GW HOMO energy ZPR in 
this work,
we compare their values to 
the ZPR
of the ionization potential (IP) as computed by the equation-of-motion ionization potential coupled cluster method with single and double excitations (EOM-IP-CCSD)~\cite{Bartlett2007}. For each of the molecules in
Thiel's set, we compute the lowest IP within EOM-IP-CCSD, corresponding to the negative of the HOMO energy according to the ionization potential theorem. We perform all EOM calculations within the Massively Parallel Quantum Chemistry (MPQC) code~\cite{Lewis2016,Peng2016}, employing the aug-cc-pVTZ basis
set as in the case of our DFT and $GW$
calculations. 


\subsection{Extracting the HOMO energy ZPR}
\label{ZPR}

\begin{figure}[tb]
    \centering
    \includegraphics[width=0.6\linewidth]{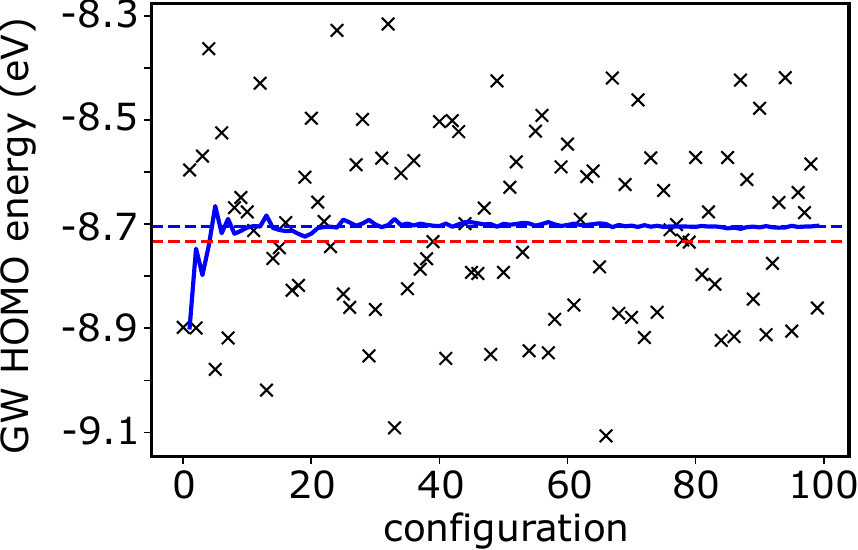}
    \caption{HOMO energies (black crosses) of furan (C$_4$H$_4$O) at different $0$\,K configurations, computed with one-shot $GW@\text{PBE0}$. The red dashed line indicates the HOMO energy in the absence of molecular vibrations, the blue dashed line is its converged $0$\,K value when vibrations are accounted for, and the blue solid line represents the cumulative average of the vibrational average of the HOMO energy.}
    \label{fig:convergence}
\end{figure}

We now combine the vibrational and electronic structure calculations of Sections\,\ref{vibrational_averages} and\,\ref{electronic_structure}
to compute the HOMO energy ZPR with different
levels of electronic structure theory. For each molecule within the Thiel's set, we generate $N=100$ displaced configurations $\mathbf{u}_i$, and following eq.\,\ref{eq:har_band_gap}, we obtain the
HOMO energy at $0$\,K as
\begin{equation}
    \label{eq:HOMO_0K}
    \epsilon_{\text{HOMO}}(0\,\mathrm{K})_{\text{MC}}=\frac{1}{100}\sum_{i=1}^{100}\epsilon_{\text{HOMO}}(\mathbf{u}_i).
\end{equation}
The values of $\epsilon_{\text{HOMO}}(\mathbf{u}_i)$ are computed within the different levels of electronic structure theory outlined
in Section\,\ref{electronic_structure}. We find that $N=100$ is generally sufficient to converge these vibrational averages,
with values in the proximity of $N=50$ already being converged within $1$\,meV in most cases. In Fig.\,\ref{fig:convergence} we show an
example of the convergence of the HOMO energy of furan at $0$\,K with $GW@\text{PBE0}$ (one-shot $GW$ with PBE0 eigenenergies and eigenstates as starting point). 

The ZPR for the HOMO energy of each molecule is then obtained as
\begin{equation}
    \label{eq:ZPR}
    \text{ZPR}=\epsilon_{\text{HOMO}}(0\,\mathrm{K})-\epsilon_{\text{HOMO}}(\mathbf{u}=\mathbf{0}).
\end{equation}
We note that for certain
limited cases of molecules, and depending on the employed DFT functional, there can be accidental
near-degeneracies between HOMO and HOMO-1, and the displacements $\mathbf{u}_i$ can sometimes induce
a change in the ordering of these orbitals. In order
to reliably apply eq.\,\ref{eq:ZPR}, one needs to
correctly identify the orbital that corresponds
to the original HOMO, which we achieve by monitoring
the expectation value of the kinetic energy operator
of the near-degenerate orbitals. This case of accidental near-degeneracies is different to the
symmetry-imposed degeneracy of molecules such as
benzene, where the HOMO and HOMO-1 are degenerate. 
In these latter degenerate cases, the ZPR of the HOMO and HOMO-1 is taken to be the average of the
two, which is a gauge-invariant quantity. 

\section{Results and discussion}
\label{Results}

\begin{figure}[tb]
    \centering
    \includegraphics[width=0.6\linewidth]{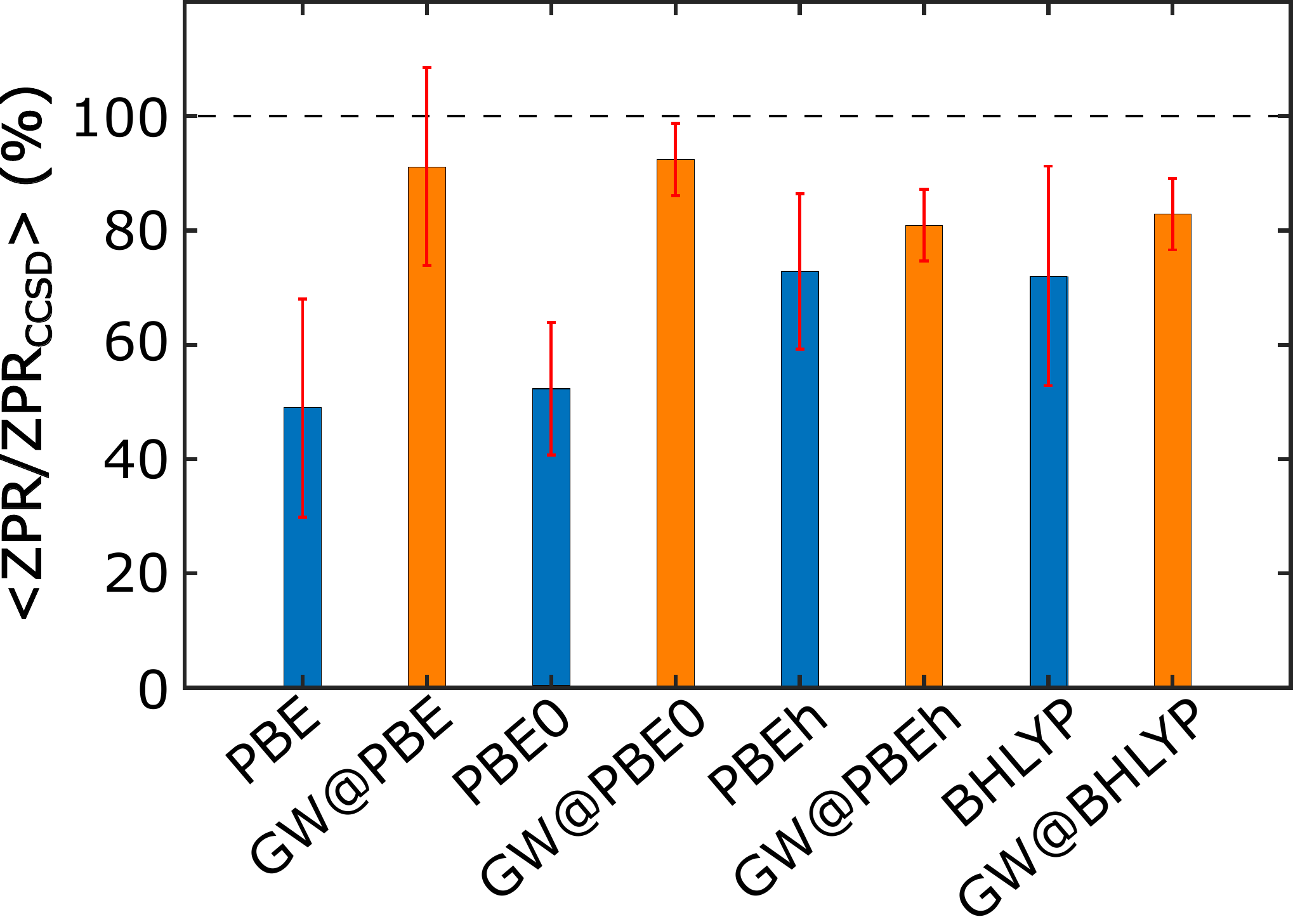}
    \caption{
    Average ratio ($\%$) of the HOMO energy ZPR obtained at the different levels of theory compared with that obtained with EOM-IP-CCSD, across the entire Thiel's set with the exception of butadiene, cyclopentadiene and naphthalene (see main text). Blue bars indicate the average ratio of the DFT HOMO energy ZPR values to EOM-IP-CCSD, while orange bars correspond to the respective $GW$ values. The red bars are the standard error to the average values. Values near $100$\% correspond to good agreement with EOM-IP-CCSD.
    }
    \label{fig:average_comparison}
\end{figure}

In Fig.\,\ref{fig:average_comparison} we show
the average (across the Thiel's set) of the ratio of
the HOMO energy ZPR at the various DFT and $GW$ levels of theory
to the ZPR obtained within EOM-IP-CCSD ($\text{ZPR}_{\text{CCSD}}$). Table\,\ref{table:mean_errors} gives the
mean absolute error (MAE) and mean signed error (MSE) of the ZPR at the different levels of theory
with respect to our EOM-IP-CCSD reference values.
All ZPR values for the HOMO energies of individual molecules are given in the supplementary material. We have excluded butadiene, cyclopentadiene and naphthalene from the averages of Fig.\,\ref{fig:average_comparison}, due to the fact that their EOM-IP-CCSD ZPR value is smaller than 2 meV, causing unphysically large oscillations in the values of the ratio presented in Fig.\,\ref{fig:average_comparison}. 

\begin{table*}[tb]
\centering
  \setlength{\tabcolsep}{8pt} 
\begin{tabular}{ccc}
\hline
 Level of theory & MSE & MAE \\
\hline
PBE & $-26.9$ & $32.8$ \\
$GW@\text{PBE}$ & $2.9$ & $11.1$\\
PBE0 & $-18.7$ & $18.8$ \\
$GW@\text{PBE0}$ & $-0.1$ & $6.9$\\
PBEh & $-5.5$ & $14.0$ \\
$GW@\text{PBEh}$ & $-5.9$ & $8.9$\\
BHLYP & $-7.8$ & $20.3$ \\
$GW@\text{BHLYP}$ & $-5.0$ & $8.6$\\
\hline
\end{tabular}
\caption{Mean signed error (MSE) and mean absolute error (MAE) for the ZPR values at the different DFT and $GW$ levels of electronic structure theory with respect to the ZPR values of our EOM-IP-CCSD reference. All values are in meV.}
\label{table:mean_errors}
\end{table*}

We first comment on the average ratio of the ZPR
obtained using DFT with the various functionals to
$\text{ZPR}_{\text{CCSD}}$, indicated with blue
bars in Fig.\,\ref{fig:average_comparison}. On average, as the content of exact exchange included in
the functional is increased, the magnitude of the ZPR increases,
approaching the EOM-IP-CCSD values. This is consistent with
reports in the literature
for the fullerene $\text{C}_{60}$~\cite{LaflammeJanssen2010}, where employing a hybrid
functional increases the magnitude of electron-phonon interactions. 
Our result here shows that
such a trend seems to hold generally across diverse
organic molecules. 
This trend of increased ZPR with
greater contents of exact exchange
is consistent with the well-known
increase in the electronic
localization in Hartree-Fock compared to semi-local DFT~\cite{Hait2018}. The inclusion of exact exchange causes increased localization between bonded
atoms, consequently resulting
in greater variation of this density
upon displacement of atoms, for
example through carbon-carbon
stretching motions. Such high-frequency motions dominate in small organic molecules
such as the ones studied here, and are indeed known to have a larger effect on electronic states with a more localized electronic density~\cite{Hele2021,Alvertis2020}. 
However, we note the inclusion of exact exchange is not sufficient to yield good agreement with the EOM-IP-CCSD
ZPR values on average. 

We now comment on the effect of
performing $GW$ calculations on top of the different
DFT starting points. The average $GW$ HOMO energy ZPR ratios to 
$\text{ZPR}_{\text{CCSD}}$ in Fig.\,\ref{fig:average_comparison} (orange bars) show
an increase compared to their DFT counterparts in
each case. This effect is particularly prominent
in the cases where functionals with a low exchange content (PBE and PBE0) are employed, with the average ZPR ratio increasing by a factor of $1.7-1.8$ in these two cases compared to DFT. This strong systematic increase of the electron-phonon interactions predicted by $GW$ compared to that computed from DFT functionals with low exchange content is consistent with multiple reports in the literature for diverse systems~\cite{Faber2011,Yin2013,Antonius2014,Monserrat2016,Li2019}. 
In fact, Refs.~\cite{Faber2011,Yin2013,Antonius2014,Monserrat2016,Li2019} all employ DFT starting points without any exact exchange. We also see from Fig.\,\ref{fig:average_comparison} that the increase of
the HOMO energy ZPR from DFT to $GW$ persists even
in cases with $50\%$ exact exchange (PBEh and BHLYP), although it
is a much weaker effect compared to using functionals
with a small content of exchange. The increase of electron-phonon interactions upon inclusion of exact exchange can in part reproduce the increase computed with $GW$. We note that the $GW$ HOMO energy ZPR values have 
significantly smaller standard errors relative to DFT (with the exception of $GW@\text{PBE}$), highlighting the overall better performance of $GW$, even over hybrid DFT calculations with substantial exact exchange included. 
The additional electronic correlations included in EOM-IP-CCSD also
lead to a small increase over
$GW$. Notably, $GW$ calculations on top of DFT starting points with no/low contents of exact exchange (PBE and PBE0 in this case)
perform best in terms of reproducing
the EOM-IP-CCSD reference.

\begin{figure}[tb]
    \centering
    \includegraphics[width=\linewidth]{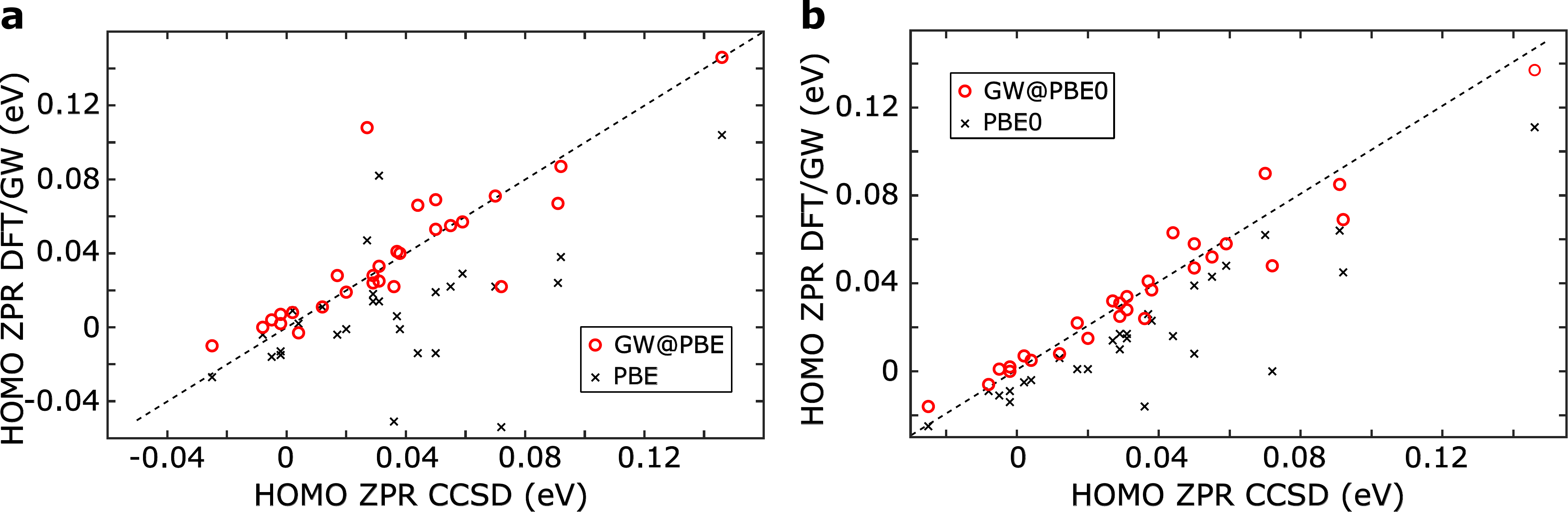}
    \caption{Comparison of the HOMO ZPR of the Thiel's set molecules computed with PBE/$GW@\text{PBE}$ (panel\,\textbf{a}) and PBE0/$GW@\text{PBE0}$ (panel\,\textbf{b}) to EOM-IP-CCSD values. Perfect agreement is indicated by the diagonal line (black dashes).
    }
    \label{fig:correlation_pbe}
\end{figure}

In order to gain a better understanding of the ZPR
averages of Fig.\,\ref{fig:average_comparison} and
the large increase of the ZPR when performing $GW$ calculations using PBE and PBE0 DFT starting points,
we plot in Fig.\,\ref{fig:correlation_pbe}
the HOMO energy ZPR with PBE and $GW@\text{PBE}$ (panel\,\textbf{a}) and with PBE0 and $GW@\text{PBE0}$ (panel\,\textbf{b}), against the EOM-IP-CCSD ZPR for each of the Thiel's set molecules. It is evident 
that $GW$ HOMO energy ZPR values are in better agreement with
EOM-IP-CCSD, and are generally larger in magnitude
compared DFT. Additionally, the $GW$ ZPR values are more
closely distributed around the black dashed line along the diagonal, indicating full agreement with EOM-IP-CCSD  values, especially in the $GW@\text{PBE0}$ case. 

\begin{figure}[tb]
    \centering
    \includegraphics[width=0.6\linewidth]{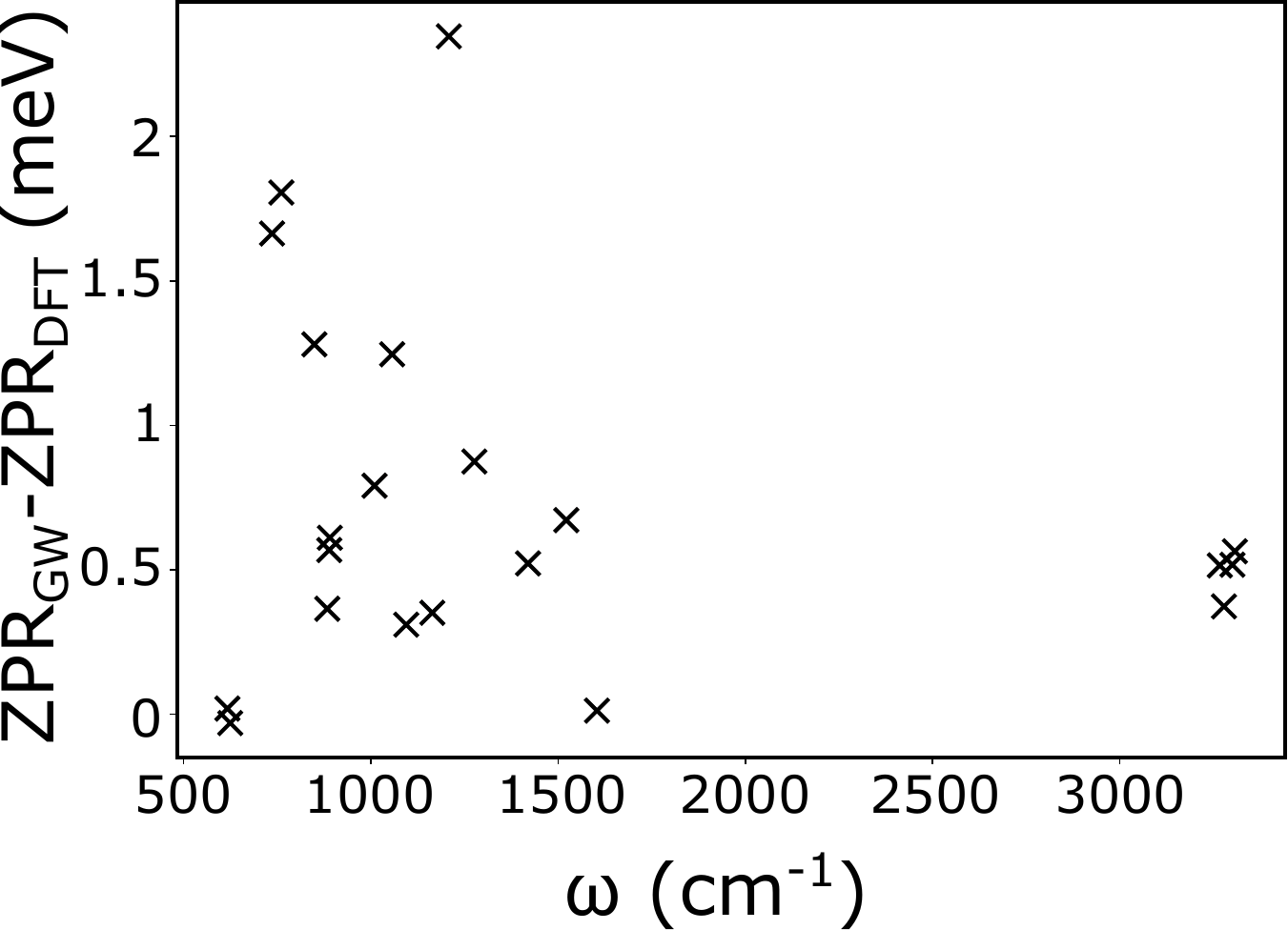}
    \caption{Vibrational mode-resolved difference in the ZPR computed with DFT PBE0 and $GW@\text{PBE0}$ for furan.}
    \label{fig:furan_dZPR}
\end{figure}

We also identify the effect
of individual vibrational modes on the ZPR differences at the different levels of theory. In Fig.\,\ref{fig:furan_dZPR} we show results for
furan, where by using the quadratic approximation of Section\,\ref{vibrational_averages} to the ZPR,
we plot the difference between the PBE0 and $GW@\text{PBE0}$ HOMO energy ZPR values decomposed into the
different normal mode contributions. Furan is representative of a more general phenomenon,
where 
every individual mode tends to contribute increased ZPR at the $GW$ level of theory compared to DFT. This is in agreement with the previous finding in the case of $\text{C}_{60}$~\cite{LaflammeJanssen2010}, where employing a hybrid functional
caused an increase of the electron-phonon coupling constants for the majority of the vibrational modes of the molecule. 

\section*{Conclusions}
\label{conclusions}
We have presented a systematic study of the zero-point renormalization of the HOMO energy of the
Thiel's set of organic molecules computed at different levels
of theory, in particular using DFT functionals with
varying degree of exact exchange, $GW$ calculations using these DFT calculations as starting points, and coupled-cluster calculations within the equation of motion formalism. We find that DFT HOMO energy
ZPR values are systematically underestimated compared to coupled-cluster values, with the
inclusion of exact exchange somewhat improving
agreement. HOMO energy ZPR values obtained within the $GW$ formalism greatly increase the electron-phonon interactions predicted by the DFT starting points
and result in much better agreement with coupled-cluster. The underestimation of the DFT ZPR is greatest when
employing functionals with low fractions of exact exchange, in agreement with examples that have been
reported in the literature over the past decade. 
Our study establishes such an increase in the magnitude of electron-phonon coupling to be a general feature of the $GW$ method for molecules, and through
comparison to coupled-cluster calculations emphasizes
that indeed computing electron-phonon interactions within methods that incorporate electronic correlations beyond DFT is a necessary step towards achieving predictive accuracy for these phenomena. Importantly, the favorable scaling of $GW$ calculations compared to coupled cluster demonstrates that \emph{ab initio} $GW$ methods can be more affordable and yet accurate for modeling electron-phonon interactions. Our results may have implications beyond molecular physics and for a wide range of systems in condensed matter, where electron-phonon interactions play an important role, including, but not limited to, molecular crystals and materials with potential
for high-temperature superconductivity. 

\section*{Acknowledgments}
This work was undertaken as part of the Photosynthetic Light Harvesting program at LBNL and primarily supported by the U.S. Department of Energy, Basic Energy Sciences, Chemical Sciences, Geosciences, and Biosciences Division, under Contract No. DE-AC02-05CH11231.
DWY was supported by the Center for Scalable Predictive methods for
Excitations and Correlated phenomena (SPEC), which is funded by the U.S. Department of Energy (DoE), Office of Science, Office of Basic Energy Sciences, Division of Chemical Sciences, Geosciences and Biosciences as part of the Computational Chemical Sciences (CCS) program at Lawrence Berkeley National Laboratory under FWP 12553. This research used resources of the National Energy Research Scientific Computing Center (NERSC), a Department of Energy Office of Science User Facility using NERSC award BES-ERCAP0023385.
Part of this work was performed using HPC resources from GENCI–TGCC (Grant 2023-gen6018).

\clearpage

\begin{figure}[tb]
    \centering
    \includegraphics[width=0.6\linewidth]{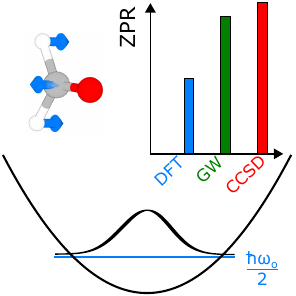}
    \caption{Table of contents figure.}
    \label{fig:toc}
\end{figure}

\textbf{}

\end{document}


\maketitle

\noindent

\clearpage

\begin{figure}[tb]
    \centering
    \includegraphics[width=0.6\linewidth]{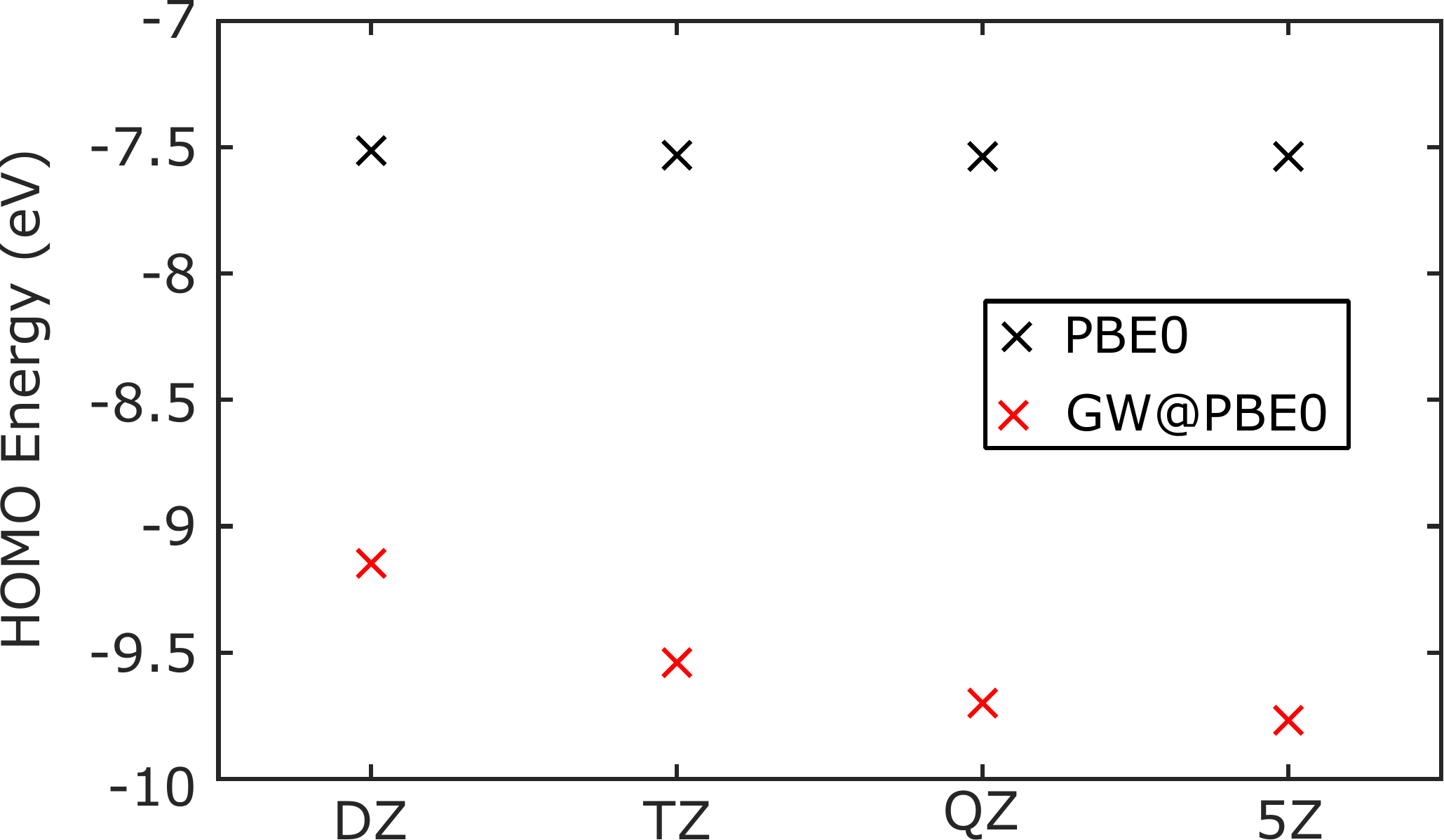}
    \caption{Convergence of the HOMO energy of pyrimidine within DFT-PBE0 and $GW@$PBE0, with respect to the size of the basis set aug-cc-pVX, with X assuming the values of the x-axis.}
    \label{fig:convergence_HOMO}
\end{figure}

\begin{figure}[tb]
    \centering
    \includegraphics[width=0.6\linewidth]{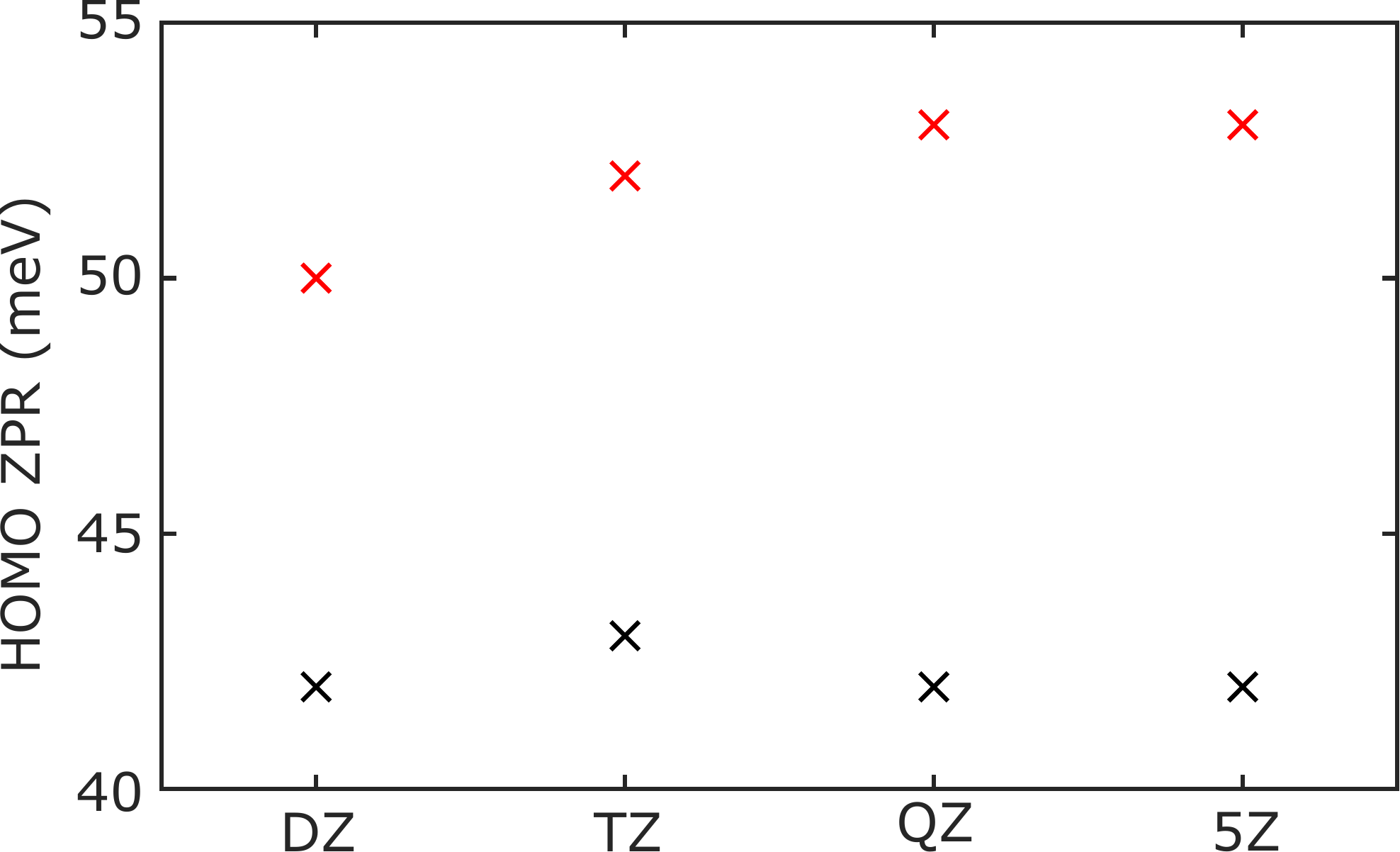}
    \caption{Convergence of the HOMO ZPR of pyrimidine within DFT-PBE0 and $GW@$PBE0, with respect to the size of the basis set aug-cc-pVX, with X assuming the values of the x-axis.}
    \label{fig:convergence_ZPR}
\end{figure}

\clearpage

\begin{table}[!ht]
    \centering
    \begin{tabular}{|l|l|l|l|l|}
    \hline
        \textbf{system} & \textbf{PBE} & \textbf{PBE0} & \textbf{PBEh} & \textbf{BHLYP} \\ \hline
        acetamide & -0.014 & 0.016 & 0.036 & 0.035 \\ \hline
        acetone & -0.014 & 0.008 & 0.031 & 0.026 \\ \hline
        adenine & 0.002 & -0.004 & -0.007 & -0.01 \\ \hline
        benzene & -0.004 & -0.009 & -0.014 & -0.018 \\ \hline
        benzoquinone & 0.038 & 0.045 & 0.143 & 0.204 \\ \hline
        butadiene & 0.009 & -0.005 & -0.002 & -0.005 \\ \hline
        cyclopentadiene & -0.015 & -0.014 & -0.012 & -0.016 \\ \hline
        cyclopropene & 0.104 & 0.111 & 0.115 & 0.105 \\ \hline
        cytosine & 0.082 & 0.017 & 0.014 & 0.012 \\ \hline
        ethene & 0.011 & 0.006 & 0.002 & -0.003 \\ \hline
        formaldehyde & 0.029 & 0.048 & 0.067 & 0.067 \\ \hline
        formamide & -0.054 & 0 & 0.059 & 0.064 \\ \hline
        furan & 0.014 & 0.015 & 0.016 & 0.015 \\ \hline
        hexatriene & -0.016 & -0.011 & -0.006 & -0.011 \\ \hline
        imidazole & 0.014 & 0.01 & 0.008 & 0.005 \\ \hline
        naphthalene & -0.013 & -0.009 & -0.006 & -0.014 \\ \hline
        norbornadiene & -0.004 & 0.001 & 0.007 & 0.003 \\ \hline
        octatetraene & -0.027 & -0.025 & -0.021 & -0.027 \\ \hline
        propanamide & -0.051 & -0.016 & 0.026 & 0.028 \\ \hline
        pyrazine & 0.024 & 0.064 & 0.064 & 0.009 \\ \hline
        pyridazine & -0.001 & 0.023 & 0.056 & 0.058 \\ \hline
        pyridine & 0.022 & 0.062 & 0.063 & 0.056 \\ \hline
        pyrimidine & 0.022 & 0.043 & 0.073 & 0.068 \\ \hline
        pyrrole & -0.001 & 0.001 & 0.004 & 0.001 \\ \hline
        tetrazine & 0.006 & 0.026 & 0.041 & 0.043 \\ \hline
        thymine & 0.018 & 0.017 & 0.016 & 0.015 \\ \hline
        triazine & 0.019 & 0.039 & 0.065 & 0.067 \\ \hline
        uracil & 0.047 & 0.014 & 0.007 & 0.006 \\ \hline
    \end{tabular}
    \caption{Zero-point renormalization of the energy of the highest occupied molecular orbital of the Thiel set molecules within density functional theory and employing the different functionals given here. All values are in eV.}
\end{table}

\begin{table}[!ht]
    \centering
    \begin{tabular}{|l|l|l|l|l|l|}
    \hline
        \textbf{system} & $\mathbf{GW}$\textbf{@PBE} & $\mathbf{GW}$\textbf{@PBE0} & $\mathbf{GW}$\textbf{@PBEh} & $\mathbf{GW}$\textbf{@BHLYP} & \textbf{IP-EOM-CCSD} \\ \hline
        acetamide & 0.066 & 0.063 & 0.044 & 0.038 & 0.044 \\ \hline
        acetone & 0.069 & 0.058 & 0.053 & 0.049 & 0.05 \\ \hline
        adenine & -0.003 & 0.005 & 0.001 & 0.0013 & 0.004 \\ \hline
        benzene & 0 & -0.006 & -0.01 & -0.01 & -0.008 \\ \hline
        benzoquinone & 0.087 & 0.069 & 0.057 & 0.053 & 0.092 \\ \hline
        butadiene & 0.008 & 0.007 & 0.004 & 0.005 & 0.002 \\ \hline
        cyclopentadiene & 0.007 & 0.002 & -0.001 & -0.002 & -0.002 \\ \hline
        cyclopropene & 0.146 & 0.137 & 0.131 & 0.128 & 0.146 \\ \hline
        cytosine & 0.025 & 0.034 & 0.032 & 0.033 & 0.031 \\ \hline
        ethene & 0.011 & 0.008 & 0.005 & 0.006 & 0.012 \\ \hline
        formaldehyde & 0.057 & 0.058 & 0.059 & 0.058 & 0.059 \\ \hline
        formamide & 0.022 & 0.048 & 0.069 & 0.066 & 0.072 \\ \hline
        furan & 0.033 & 0.028 & 0.026 & 0.028 & 0.031 \\ \hline
        hexatriene & 0.004 & 0.001 & -0.001 & -0.002 & -0.005 \\ \hline
        imidazole & 0.028 & 0.025 & 0.02 & 0.022 & 0.029 \\ \hline
        naphthalene & 0.002 & 0 & -0.002 & -0.007 & -0.002 \\ \hline
        norbornadiene & 0.028 & 0.022 & 0.018 & 0.018 & 0.017 \\ \hline
        octatetraene & -0.01 & -0.016 & -0.018 & -0.021 & -0.025* \\ \hline
        propanamide & 0.022 & 0.024 & 0.021 & 0.016 & 0.036 \\ \hline
        pyrazine & 0.067 & 0.085 & 0.011 & 0.107 & 0.091 \\ \hline
        pyridazine & 0.04 & 0.037 & 0.04 & 0.033 & 0.038 \\ \hline
        pyridine & 0.071 & 0.09 & 0.087 & 0.088 & 0.07 \\ \hline
        pyrimidine & 0.055 & 0.052 & 0.034 & -0.004 & 0.055 \\ \hline
        pyrrole & 0.019 & 0.015 & 0.013 & 0.014 & 0.02 \\ \hline
        tetrazine & 0.041 & 0.041 & 0.041 & 0.04 & 0.037 \\ \hline
        thymine & 0.024 & 0.031 & 0.026 & 0.03 & 0.029 \\ \hline
        triazine & 0.053 & 0.047 & 0.051 & 0.047 & 0.05 \\ \hline
        uracil & 0.108 & 0.032 & 0.025 & 0.027 & 0.027 \\ \hline
    \end{tabular}
    \caption{Zero-point renormalization of the energy of the highest occupied molecular orbital of the Thiel set molecules within coupled cluster (IP-EOM-CCSD) and the $GW$ approximation using a starting point of density functional theory with the different functionals given here. All values are in eV.
    *The IP-EOM-CCSD values for octatetraene have been obtained within the smaller aug-cc-pVDZ basis set, due to memory issues encountered when employing aug-cc-pVTZ.}
\end{table}
